\documentclass[prl]{revtex4}
\usepackage{graphicx}
\usepackage{amsmath}
\usepackage{amssymb}
\usepackage{color}
\usepackage[normalem]{ulem}
\usepackage{epsfig}
\usepackage{amssymb,amsfonts,amsmath,color}

\newcommand{\stkout}[1]{\ifmmode\text{\sout{\ensuremath{#1}}}\else\sout{#1}\fi}

\begin{document}

\title{Phase diagram for logistic systems under bounded  stochasticity}

\author{Yitzhak Yahalom and Nadav M. Shnerb}

\affiliation{Department of Physics, Bar-Ilan University,
Ramat-Gan IL52900, Israel.}

\begin{abstract}
\noindent  Extinction is the ultimate absorbing state of any stochastic birth-death process, hence the  time to extinction is an important characteristic of any natural population. Here we consider logistic and logistic-like systems under the combined effect of demographic and bounded environmental stochasticity. Three phases are identified: an inactive phase where the mean time to extinction $T$ increases logarithmically with the initial population size, an active phase where $T$ grows exponentially with the carrying capacity $N$, and temporal Griffiths  phase, with power-law relationship between $T$ and $N$. The system supports an exponential phase only when the noise is bounded, in which case the continuum (diffusion) approximation breaks down within the Griffiths phase. This breakdown is associated with a crossover between qualitatively different survival statistics and decline modes. To study the power-law phase we present a new WKB scheme which is applicable both in the diffusive and in the non-diffusive regime.
\end{abstract}

\maketitle

Noise and fluctuations are ubiquitous features of living systems. In particular, the reproductive success of individuals is affected by many random factors. Some of these factors, like the local density of nutrients or accidental encounter with predators, act on the level of a single individual. Others, like fluctuations in temperature and precipitation rates, affect many individuals coherently. The corresponding theory distinguishes between \emph{demographic stochasticity} (shot noise), i.e., those aspects of noise that influence individuals in an uncorrelated manner, and \emph{environmental stochasticity}, that acts on entire populations~\cite{lande2003stochastic,ovaskainen2010stochastic}.

For a population of size $n$, demographic noise yields ${\cal O} (\sqrt{n})$  abundance fluctuations while environmental stochasticity leads to ${\cal O} (n)$  variations.  Accordingly, for large populations environmental stochasticity is the dominant mechanism. A few recent analyses of empirical studies confirm this prediction \cite{leigh2007neutral,kalyuzhny2014niche,kalyuzhny2014temporal,
chisholm2014temporal}. However, the demographic noise controls the low-density states and must be taken into account for calculations of extinction times or fixation probabilities.  Consequently, the study  of models that combine deterministic effects,  temporal environmental stochasticity and demographic noise,  received a considerable  attention during the last years ~\cite{kessler2014neutral,saether2015concept,cvijovic2015fate,
danino2016stability,
fung2016reproducing,hidalgo2017species,wienand2017evolution}.

Almost any model of population dynamics includes two basic ingredients, exponential grows and resource competition.  In particular,  in the famous logistic equation,
\begin{equation} \label{eq1}
\frac{dn}{dt} = r_0 n - \beta n^2,
\end{equation}
$r_0$ is the basic reproductive number (low-density growth rate) and the  $\beta$ term reflects a density-dependent crowding effect, so the per-capita growth rate declines linearly with $n$.

 A wide variety of similar models include the $\theta$-logistic  equation (where the growth rate declines like $n^\theta$),  ceiling models (growth rate is kept fixed but the population cannot grow above a given carrying capacity), Ricker dynamics and so on. All these models support a transcritical bifurcation at $r_0=0$: when $r_0<0$ the extinction point $n=0$ is stable, while for $r_0>0$ it becomes unstable and the system admits a finite population stable state at $n^*$ [e.g., $n^*=r_0/\beta$ for the logistic equation (\ref{eq1})].

Since the actual number of individuals in a population is always an integer,  Equations like  (\ref{eq1}) can only be interpreted as the deterministic limit of an underlying stochastic process. For any process with demographic noise the empty state $n=0$ is the only absorbing state, so each population must reach extinction in the long run. Under purely demographic noise the bifurcation point separates two qualitatively different behaviors of the mean time to extinction $T$.  When $r_0<0$  the extinction time is logarithmic in the initial population size, while for $r_0>0$ the time to extinction grows exponentially with  $n^*$~\cite{elgart2004rare,assaf2006spectral,kessler2007extinction,ovaskainen2010stochastic}.

  To understand the lifetime of empirical populations one would like to study a logistic system under the influence of  both demographic and environmental stochasticity. This problem was considered by a few authors~\cite{lande2003stochastic,kamenev2008colored,spanio2017impact,wada2018extinction} for the case where the strength of the environmental fluctuations is \emph{unbounded}, e.g., when the state of the environment undergoes an Ornstein-Uhlenbeck process. In such a case there are always rare periods of time in which the net growth rate is negative, and (as we shell see below) these periods dominate the asymptotic behavior of the extinction times. As a result, the system admits only two phases: an inactive (logarithmic) phase for $r_0<0$  and a temporal Griffiths phases~\cite{vazquez2011temporal}, where $T$ scales like a power-law with $n^*$, for $r_0>0$.

\begin{figure}
\includegraphics[width=8cm]{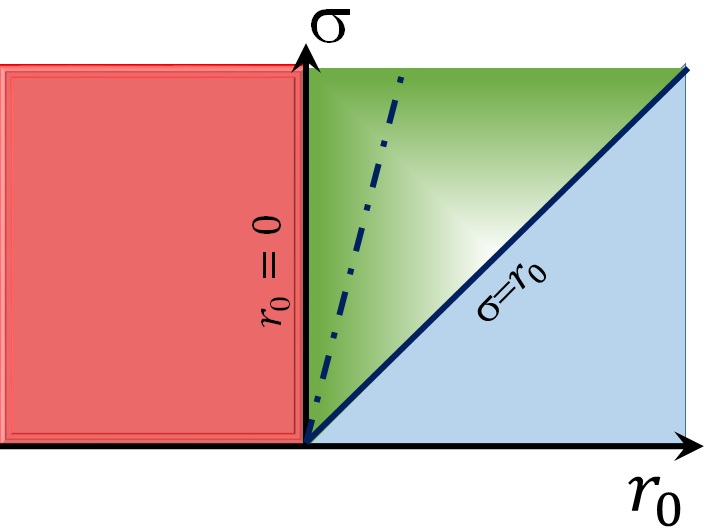}
\caption{A phase diagram for a logistic system under  bounded  stochasticity, presented in the $r_0$-$\sigma$ plane. In the inactive phase  ($r_0  <0$, red) the time to extinction scales like $\ln n$ where $N$ plays no role. In the active  phase  $r_0>\sigma$ (blue) the extinction time grows exponentially with $N$. Under pure demographic noise (along the $\sigma=0$ axis) the transition occurs at $r_0=0$. When $\sigma>0$ the logarithmic and the exponential phases are separated by a finite power-law region (temporal Griffiths phase, green).  The dotted line indicates the failure of the continuum (diffusive) approximation and the crossover from soft to sharp decline. }\label{fig1}
\end{figure}

Here we would like to consider another scenario, a system under \emph{bounded} environmental variations.  Since the noise is bounded, for large enough $r_0$ the growth rate is always positive, so the system allows for three phases: logarithmic, power-law (temporal Griffiths phase) and exponential (see Figure \ref{fig1}). This insight allows one to identify the failure of the standard analytic tool, the diffusion (continuum) approximation, inside the temporal Griffiths phase. To overcome that we  provide an alternative WKB analysis which is valid all over the power-law region. Our analysis reveals a crossover between two, qualitatively different, extinction dynamics. This distinction, in turn, may be relevant to several key concepts in the modern theory of viability and coexistence~\cite{schreiber2012persistence,ellner2018expanded,barabas2018chesson}. These connections will be expanded upon towards the end of the paper.

Our model system, chosen to facilitate the numerical calculations, involves two "species" (types, alleles) competition in a finite community with one-sided mutation~\cite{karlin1981second}. In this system the number of individuals, $N$, is always fixed, where $n$ of them belong to species A and $N-n$ to species B. At each elementary step two individuals are chosen at random for a duel, the loser dies and the winner produces a single offspring~\cite{maritan1}. The possible outcomes of the duels (expressions above arrows represent probabilities) are,
\begin{eqnarray} \label{eq1a}
 B+B \xrightarrow{1} 2B \quad A&+&A  \xrightarrow{1-\nu} 2A  \quad A+A  \xrightarrow{\nu} A+B \\ \nonumber
 A+B \xrightarrow{1-P_{A}} 2B \quad A&+&B \xrightarrow{P_{A}(1-\nu)} 2A \quad A+B \xrightarrow{\nu P_{A}} A+B,
 \end{eqnarray}
 where $\nu$ is the chance of a mutation event, in which the offspring of an $A$ is a $B$.

 An $A$ individual wins an interspecific duel with probability $P_A = 1/2+s(t)/4$, where $s(t) = s_0 + \eta(t)$ and $\eta(t)$ is a zero-mean random process. Following~\cite{hidalgo2017species}  we consider a system with dichotomous (telegraphic) environmental noise, so $\eta = \pm \sigma$ (see Supplementary Material, section II~\cite{note1}). After each elementary step $\eta$ may switch (from $\pm \sigma$ to $\mp \sigma$) with probability $1/N\tau$, so the persistence time of the environment is taken from a geometric distribution with mean $\tau$ generations, where a generation is defined as $N$ elementary duels.

As required, this process supports an absorbing attractive fixed point at $n^*=0$ when $r \equiv  s(t)-\nu <0$ (more accurately the condition is $\tilde{s} \equiv s(1-\nu/2)<\nu$. In what follows we neglect this tiny factor and use $s$ for $\tilde{s}$) and an active attractive fixed point at $n^*=N[1-\nu/s(t)]$ when $r>0$.

Using the procedure described in~\cite{danino2016stability}, one may derive a discrete Backward Kolmogorov equation (BKE) for this stochastic process. The BKE may be solved numerically, by inverting the corresponding matrix, to obtain $T(n)$, the mean time to extinction for a system with $n$ A-type individuals. The mean is taken over both histories and the initial state of the system (plus or minus $\sigma$). The numerical results presented below were obtained from the BKE using this technique.  For large-$N$ systems we implemented, instead of direct inversion of a matrix, a transfer matrix approach that allows us to increase the numerical accuracy.

 \emph{If} $N \gg 1$ and the diffusion approximation is applicable, $n$ may be replaced by the fraction $x=n/N$ and  $n \pm 1$ by $x \pm 1/N$. Expanding all the relevant quantities to second order in  $1/N$, and using the dominant balance analysis presented in~\cite{danino2016stability}, it can be shown that  $T(x)$ satisfies,
 \begin{eqnarray} \label{eq9}
\left(s_0 - \frac{\nu}{1-x} + g (1-2x) \right) \frac{\partial T(x)}{\partial x} &+& \left(\frac{1}{N} + g x(1-x) \right) \frac{\partial^2 T(x)}{\partial x^2} = -\frac{1}{x(1-x)} \nonumber \\ T(0) &=& 0 \qquad \left.  \frac{\partial T(x)}{\partial x} \right|_{x=1}=\frac{1}{\nu},
\end{eqnarray}
where  $g \equiv \sigma^2 \tau/2$ is the diffusion constant along the log-abundance axis.  We solved Eq. (\ref{eq9}) separately in the inner region $x \ll 1$  and in the outer region $x \gg 1/Ng$, using asymptotic matching to obtain, for $1/Ng \ll x \ll 1$,
\begin{equation} \label{large}
T(x) = \left([Ng]^{r_0/g}-x^{-r_0/g} \right) \frac{\Gamma(r_0/g)}{r_0} \left( \frac{g}{\nu} \right)^{r_0/g} - \frac{\ln Ngx}{r_0}.
\end{equation}
Accordingly, the time to extinction is logarithmic in $n=Nx$ when $r_0$ is negative (red region in Figure \ref{fig1}). If $r_0$ is positive the mean lifetime, \emph{for any initial conditions},  grows like $N^{r_0/g}$, since the chance of small population (even a single individual) to grow and to reach the carrying capacity is $N$-independent.  These results are in complete agreement with former studies~\cite{lande2003stochastic,kamenev2008colored,spanio2017impact,wada2018extinction} of different logistic-like models, indicating the universality of the large $N$  behaviour for  all the systems that support a transcritical bifurcation.

However, for finite noise this continuum approximation must fail somewhere inside the power-law phase. Eq. (\ref{large}) suggests a power-law dependence of $T$ on $N$ for any $r_0>0$, but this cannot be the case for  $r_0>\sigma$ (light blue region of Fig. \ref{fig1}), where even in the pure $(-\sigma)$ state the time to extinction grows exponentially with $N$~\cite{elgart2004rare,assaf2006spectral,kessler2007extinction,ovaskainen2010stochastic} and occasional jumps to the  $+\sigma$ state can only increase  stability.

To study the system when the continuum approximation fails, we adopt a version of the WKB analyses presented and discussed in \cite{kessler2007extinction,meyer2018noise}. We shall neglect the demographic noise and replace it (as in \cite{lande2003stochastic,hidalgo2017species}) by an absorbing boundary condition at $x=1/N$.  The abundance dynamics is given by $\dot{x} = (r_0 \pm \sigma)x-\beta x^2$, where the environment stays in the same state (plus or minus $\sigma$) for $\tilde{\tau}$  generations and than switches, with probability $1/2$, to the other state (minus or plus $\sigma$).

Under this dynamics, if the system reaches $x$ at certain time $t$, then one time increment before, at  $t- \tilde{\tau}$, it was either at $x_+(x)$ or at $x_-(x)$. Equivalently one may define $y \equiv \ln x$ and $y_\pm \equiv \ln x_\pm$.  The probability to find the system at the log-density $y$ at time $t$, $P(y,t)$ satisfies the master equation,
\begin{equation} \label{master}
\frac{dP(y,t)}{dt} = \frac{1}{2} \left[-2P(y) + P(y_+) + P(y_-) \right].
\end{equation}

At long times  $P(y,t)$ converges to its quasi-steady state for which $dP/dt \approx 0$~\cite{kessler2007extinction}. Given $P(y)$, the $N$-dependence of the mean time to extinction is inversely proportional to the rate of extinction, which is  the probability to find the system with less than one individual ($0<x<1/N$), so,
\begin{equation} \label{rate}
{\rm Rate} \sim \int_{-\infty}^{-\ln N} P(y) \ dy.
\end{equation}

When $x$ is vanishingly small $x_\pm \approx x e^{-\tilde{\tau}(r_0\pm \sigma)}$. Accordingly, in the extinction zone the quasi-steady state satisfies,
\begin{equation}
P(y-\tilde{\tau}[r_0+\sigma]) + P(y-\tilde{\tau}[r_0-\sigma])=2 P(y).
\end{equation}
Instead of expanding $P(y_\pm)$ to second order in $\tilde{\tau}$ (this yields the continuum Fokker-Planck equation and the power-law  of the continuum limit) we assume that $P(y) = e^{S(y)}$ and implement the continuum approximation for $S$, replacing $S(y+\Delta y)$ by $S(y)+\Delta y S'(y)$,  so $S'(y)$ is obtained as  a solution of the transcendental equation
\begin{equation} \label{trans}
\exp \left(- \tilde{\tau}   r_0 S'\right) \cosh\left( \tilde{\tau} \sigma S' \right) = 1.
\end{equation}
This equation does not depend on y, so $S'=q$ and $S \sim q y$, where $q$ is some constant. Accordingly $P \sim \exp(q y)$ and ${\rm Rate} \sim N^{-q}$, so the time to extinction behaves like $T \sim N^q.$

In the limit $r_0 \ll \sigma$ one expects $q \ll 1$. In that case both $q \tilde{\tau} r_0$ and $q \tilde{\tau} \sigma$ are small numbers and  Eq. (\ref{trans}) yields,
\begin{equation}  \label{res1}
q = \frac{2 r_0}{(\sigma^2+r_0^2) \tilde{\tau} } \approx \frac{2 r_0}{\sigma^2 \tilde{\tau} },
\end{equation}
where the last approximation reflects a self consistency requirement for $q \tilde{\tau} r_0 \ll 1$. On the other hand if $q \tilde{\tau} \sigma$ is large,
\begin{equation} \label{res2}
q = \frac{\ln 2}{\tilde{\tau}(\sigma-r_0)}.
\end{equation}

The case (\ref{res1}) corresponds to the regime where the continuum approximation holds. In that case the typical extinction trajectory is a random walk excursion in the log-abundance space (see below).  Since the variance of $M$ random numbers, picked independently from an exponential distribution with mean $\tau$ with alternating signs, is equal to the variance of the sum of $M$ random steps of length $\tilde{\tau}$, $\tilde{\tau}=\tau$  and
\begin{equation} \label{diff}
T \sim N^{r_0/g},
\end{equation}
 in agreement with the large $N$ asymptotics of (\ref{large}).

In the other extreme (\ref{res2}) extinction occurs due to a (rare) long  sequence of bad years, so $\tilde{\tau}$ must be compared with the tail of the corresponding exponential distribution, in which case $\tilde{\tau} = \tau \ln 2$, hence in this regime
\begin{equation} \label{nondiff}
T \sim N^{1/[\tau(\sigma-r_0)]}.
\end{equation}
 This result indicates that the diffusion approximation indeed fails (the result depends on $\tau$ and $\sigma$ separately, not on $g$) and that the power diverge when $r_0 \to \sigma$, i.e., at the transition between the temporal Griffiths phase and the exponential phase.

\begin{figure}
	\includegraphics[width=8cm]{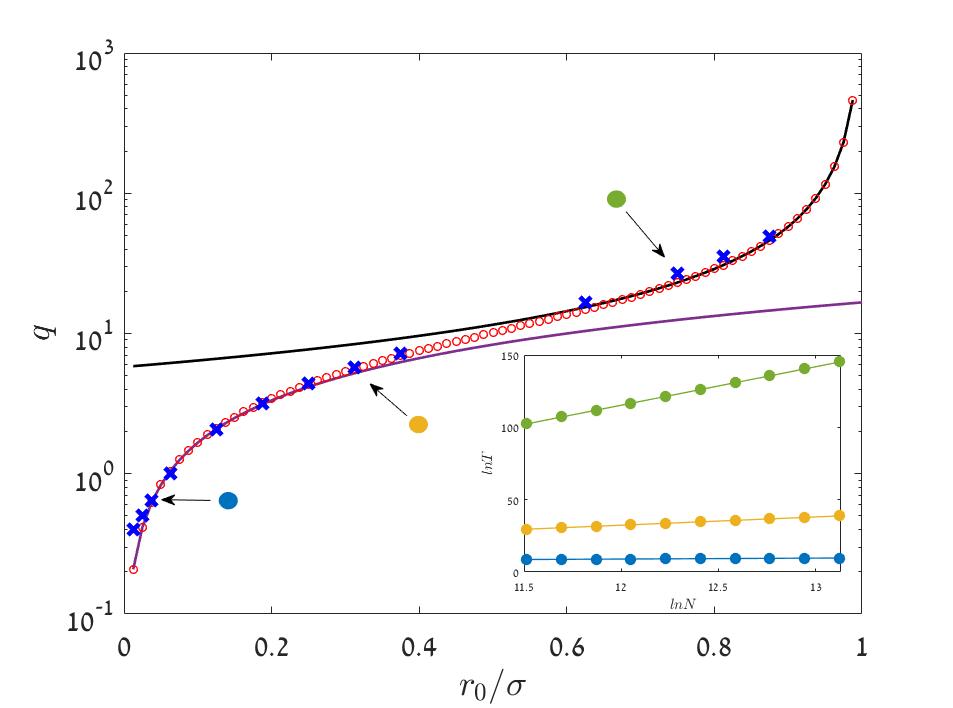}
	\caption{In the temporal  Griffiths phase $T \sim N^{q}$. The main panel shows  $q$ vs. $r_0/\sigma$ as obtained from numerical solution of Eq. (\ref{trans}) (red open circles), in comparison with the asymptotic expressions for the diffusive regime [Eq. (\ref{diff}), purple line] and in the large $r_0$ regime [Eq. (\ref{nondiff}), black line]. In the inset we present results for $T(N)$ as obtained from the numerical solution of the exact backward Kolomogorov equation for $r_0 = 0.003$ (blue circles)  $0.025$ (yellow) and $0.06$ (green). By fitting these numerical results (full lines) one obtains the actual power $q$, and the outcomes are represented by blue $X$s in the main panel (the $X$s that correspond to the three specific cases depicted in the inset are marked by arrows). In general the WKB predictions fit quite nicely the numerical outcomes, and the slight deviations in the low $r_0$ region are due to the prefactors of the power law [in these cases the numerical $T(N)$ graph fits perfectly the predictions of Eq. (\ref{large})]. All the results here were obtained for  $\sigma = 0.08, \ \tau = 3/2, \ \nu = 0.04$.   }\label{fig4}
\end{figure}

Beside these limits, The transcendental  equation  (\ref{trans}) has to be solved numerically. In figure \ref{fig4} these numerical solutions are compared with the results obtained  from a numerical solution of the BKE and with the asymptotic expressions (\ref{diff}) and (\ref{nondiff}).

As discussed in length in the Supplementary, section I~\cite{note1}, our WKB analysis provides another evidence for the universality of all logistic-like (transcritical) systems. The only features that were used to establish Eq. (\ref{trans}) are the existence of an upper bound and the linearity of the growth rate at small $x$.

In the Supplementary (section III) we also show that the qualitative features of the extinction process change along the power-law phase, together with the functional form of the survival probability function $Q(t)$ (the chance of the system to survive until $t$).

\begin{figure}
\includegraphics[width=7cm]{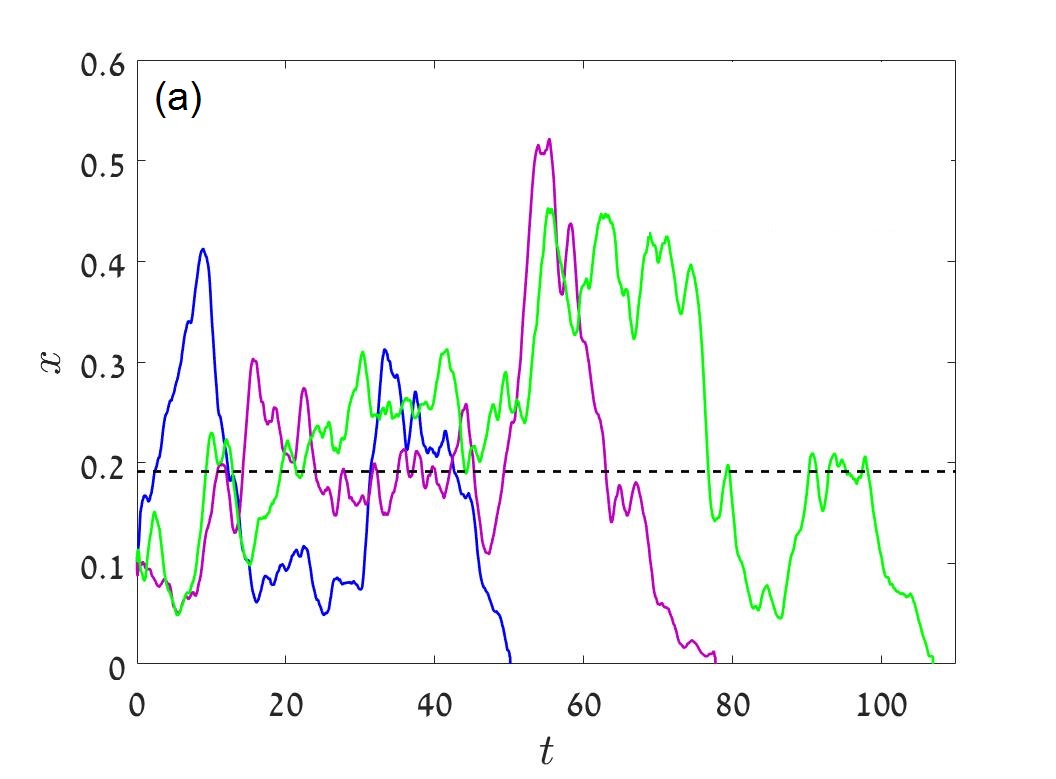}
\includegraphics[width=7cm]{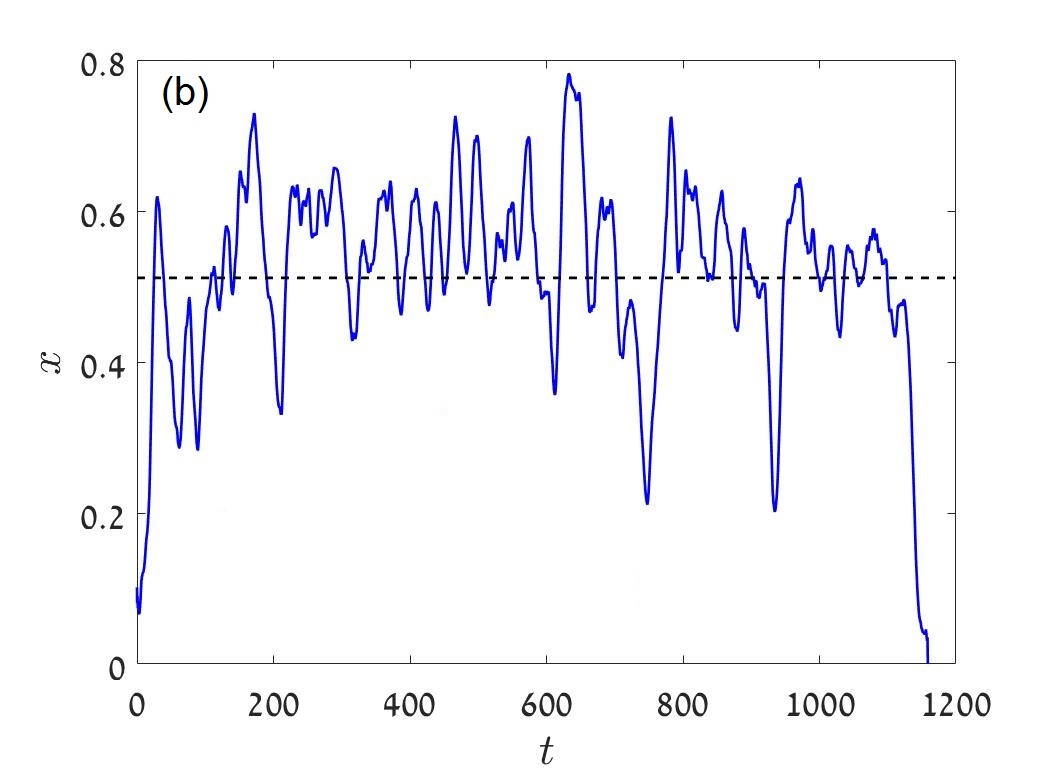}
\caption{Typical trajectories (frequency vs. time) for a system with $\tau=1$, $\sigma =0.11$ and $\nu = 0.1$, where $r_0=0.02$ (a)  and $r_0 = 0.105$ (b).  The dashed line corresponds to $x^*$, the point where the mean (over environmental conditions) growth rate is zero.  In panel (b) the population fluctuates most of its lifetime in a relatively narrow band around $x^*$, extinction happens due to the accumulation of rare sequences of bad years and the  decline time is logarithmic in $N$ (sharp decline). As $r_0$ becomes smaller (panel a) the fluctuations are comparable with $x^*$, hence the decline time becomes a finite fraction of the lifetime (soft decline).}\label{fig3}
\end{figure}

Deep inside the temporal Griffiths phase (and in the exponential phase) the system spent most of its time fluctuating around $x^*$ (the point where the mean of $\dot x$ vanishes, when the average is taken over the two signs of $\sigma$). Extinction reflects a rare event, an improbable series of bad years and/or excess deaths. Accordingly, the decline time (roughly speaking, the duration of the last excursion from $x^*$ to extinction) scales like $\ln N$~\cite{lande2003stochastic}, and is negligible with respect to the lifetime $T$ (see Fig \ref{fig3}b). In that "sharp decline" case the system has no memory: during each segment of time   either  the catastrophe occurs or not. Accordingly, $Q(t) \sim \exp(t/T)$, where $T$ is the mean time to extinction calculated above. As discussed in the supplementary, this behavior is associated with a gap in the spectrum of the corresponding Markov matrix.

In the diffusive regime, close to the extinction phase, the spectral gap closes down and the associated survival probability is $Q(t) \sim \exp(t/t_0)/t^{1/\rho}$, where $\rho$ is related to the dispersion relation of the Markov matrix  and  $T$ is proportional to  $t_0$. In that case the decline time is relatively long ("soft decline", Figure \ref{fig3}a) and an excursion to extinction is a typical first passage trajectory of a random walker along the log-abundance axis.

Our results seems to be relevant to two important issues in population and community ecology: modern coexistence theory and the assessment of population viability.

Modern coexistence theory (MCT) have gained a lot of attention in recent years~\cite{ellner2018expanded,barabas2018chesson}. In  MCT  ``coexistence" is declared if the steady state probability distribution function is normalizable~\cite{schreiber2012persistence}. For the system considered here, close to zero $P \sim e^{qy} \sim x^{q-1}$,  so the MCT persistence criteria is satisfied if $q>0$, i.e., for any $r_0>0$.

However, the main factor that determines ecological stability and  species turnover rates is the mean time to extinction. Given Eq. (\ref{rate}), one realizes that the coexistence criteria of MCT only guarantees that the time to extinction diverges with $N$, but this divergence may be as slow as $N^{\epsilon}$ for arbitrary small $\epsilon$ if $g=\epsilon r_0$. Accordingly, we believe that an instructive classification of populations stability properties must use phase diagrams like  Fig. \ref{fig1}, instead of being focused on (co)``existence". In particular, for populations in the exponential phase extinction risk is usually negligible, while in the sharp decline region extinction occurs due to rare events so our predictive ability is quite limited. On the other hand in the inactive/soft decline regions extinction risk is high and is strongly related to the observed dynamics, so one may identify risk factors (like grazing or habitat loss) and try to avoid them. 

Practically, in empirical studies of birds and plants populations  an initial abundance  $n_0$ was measured and the survival probability $Q$ was examined after a fixed time interval $t$ ~\cite{matthies2004population,jones1976short}. If $n_0$ may be taken as a proxy for the carrying capacity, the results seem to indicate that these systems are in the power-law phase (see Supplementary section IV, where the empirical results are reproduced and analyzed). However, a single observation of $n_0$ cannot provide a reliable estimation of the carrying capacity in  the soft decline regime. Large scale empirical studies of $Q(t)$ (like those presented in \cite{keitt1998dynamics,bertuzzo2011spatial}) suggest an exponentially truncated power law. If one likes to interpret these results as reflecting purely local dynamics under environmental stochasticity, it implies that the decline in these systems is indeed soft.

In spatially extended systems the correlation length of environmental fluctuations plays an important role. When the linear size of the system is much smaller than the correlation length, temporal fluctuations are global. This case was examined recently in~\cite{barghathi2017extinction}, and is expected to show similarities to the dynamics of a local population. On the other hand, when the correlation length is shorter than the population range migration tends to average out the stochastic effects so the effective strength of stochasticity decreases and $T$ increases. Such an increase  was reported by~\cite{bertuzzo2011spatial}.

We acknowledge many helpful discussions with David Kessler.  This research  was supported by the ISF-NRF Singapore joint research program (grant number 2669/17).

\bibliography{refs}

\clearpage
\widetext
\begin{center}
\Large{\bf{Supplementary information to: \\ Phase diagram for logistic systems under bounded  stochasticity.}}
\end{center}
\setcounter{equation}{0}
\setcounter{section}{0}
\setcounter{figure}{0}
\setcounter{table}{0}
\setcounter{page}{1}
\makeatletter
\renewcommand{\theequation}{S\arabic{equation}}
\renewcommand{\thefigure}{S\arabic{figure}}

\vspace{1cm}

In this supplementary we will discuss the generality of our results, and consider some features of the probability distribution function (pdf) $f(t) dt$, i.e., the chance that extinction occurs at time $t$.

\section{Universality}

The classical logistic growth equation may be written as
\begin{equation} \label{s1}
\frac{dn}{dt} = r_0 n- \beta n^2.
\end{equation}

This dynamics supports a transcritical bifurcation at $r_0=0$. For $r_0<0$, zero is a stable fixed point and there is no other fixed point in the "physical" regime $n \ge 0$ (if $n$ represents population abundance, there are no negative populations). When $r_0>0$ the extinction fixed point at zero becomes unstable and there is a stable fixed point at $n^*=r_0/\beta$. Below the transition the population decays exponentially to zero (so for  long times $n \sim \exp(-|r_0|t)$), above the transition the population eventually converges to $n^*$. At the transition point ($r_0=0$) the population still shrinks to zero but its long time decay satisfies a power law, $n \sim 1/t$.

In general, a given deterministic equation (like Eq. \ref{s1}) may be obtained as the large $N$ limit of many "microscopic" (individual based) processes.  In the main text we considered one specific example, namely two species competition with one sided mutation, a classical population-genetics problem taken from \cite{karlin1981second}. Beside its concrete importance, this system is technically tractable since it corresponds to a zero sum game so the total community size $N$ is strictly fixed and still the system shows negative density dependence.

What about other microscopic processes that yield, in their deterministic limit, a transcritical bifurcation? It is widely believed that, although some details may depend on the microscopy of the process, the main characteristics of the behavior: the different phases, the functional $N$ dependence in each phase, and the behavior at the transition points - are \emph{universal}, i.e., are independent of the microscopy. For example, in \cite{kessler2007extinction} the mean time to extinction of the  two processes $A \to A+A, \  \ A+A \to \varnothing$ and $A \to A+A, \  A \to \varnothing, \  \ A+A \to \varnothing$ was calculated. Both these processes are logistic, and under pure demographic stochsticity $T$ grows like $\exp(\alpha n^*)$, where $n^*$ is the number of individuals in the quasi-stationary state. The value of $\alpha$ does depend on the microscopy and the two different  models yield different $\alpha$s, but the exponential growth of $T$ with $n^*$ is a universal feature. Accordingly, for different microscopic models one may expect the same phase diagram with exponential, power-law and logarithmic regimes, but the prefactors and the constants may differ.

In the literature one may find other models that belong to the equivalence class of the logistic growth with environmental stochasticity. These include a model with ceiling (i.e., for which the growth rate is density-independent until it reaches a prescribed value $n^*$, where reflecting boundary conditions are imposed~\cite{lande1993risks,lande2003stochastic},  simple logistic equation~\cite{kamenev2008colored,vazquez2011temporal} and so on. Indeed for these models the authors  obtained the same $N$ dependence that we obtained here when the diffusion approximation holds (to the left of the dashed line in Fig. 1 of the main text).

We would like to stress that the WKB analysis presented in the main text allows us to suggest a much stronger statement.

As explained, the chance of extinction, and the associated timescale, are given by the behavior of $P(x)$ at $x < 1/N \ll 1$ ($ {\rm Rate} \sim \int_0^{1/N} P(x) \ dx $). This behavior depends, in turn, on the small $x$ dependency of $x_{pm}$. A solution to Eq. (8) of the main text exists only in the power-law phase (it requires $r_0>0$ and $r_0<\sigma$).  When it exists,  Eq. (8) assures that in the $x \ll 1$ ($y \ll 0$)  regime  $S=qy$  and the mean time to extinction is a power law in $N$.

 Accordingly, our WKB analysis shows that for \emph{any} microscopic model, the time to extinction is a power-law in $N$  when the following conditions are met:
 \begin{itemize}
 \item The probability distribution function $P(x)$ is normalizable.
   \item The dynamics allows for periods of growth and periods of decline.
   \item When  $x \ll 1$ the time-averaged growth rate is positive ($r_0>0$).
   \item When $x \ll 1$  the growth/decline are exponential.
 \end{itemize}
 These  features are common to any system that fluctuates below and above a transcritical bifurcation.

\clearpage

 \section{Dichotomous (telegraphic) and other types of noise}

 In the main text we have considered a special type of environmental stochasticity, in which the system flips between two states (good and bad years, say). Both white Gaussian noise and white Poisson noise can be recovered from this dichotomous (telegraphic) noise by taking suitable limits~\cite{ridolfi2011noise}, so the results obtained here are quite generic.

 As an example, if the environmental conditions are picked from a Gaussian distribution of a certain width with correlation time $\tau_1$, one may easily imitate these features by taking a dichotomous noise that flips between two values, $\pm \sigma$, with much shorter correlation time $\tau$. With the appropriate choice of $\tau$ and $\sigma$, the binomial distribution of $\sigma_{eff}$,  the average fitness between $0<t<\tau_1$,
 \begin{equation}
 \sigma_{eff} = \frac{\tau_1}{\tau} \sum_i^{\tau_1/\tau} \sigma_i,
 \end{equation}
 will correspond to the bulk properties of any required Gaussian noise, since the Gaussian distribution is the limit of a binomial distribution.

 However, while the Gaussian distribution is unbounded, the distribution of $\sigma_{eff}$ is clearly bounded; the convergence  to a Gaussian takes place in the bulk but  the tails are truncated.

\begin{figure}[h]
\includegraphics[width=8cm]{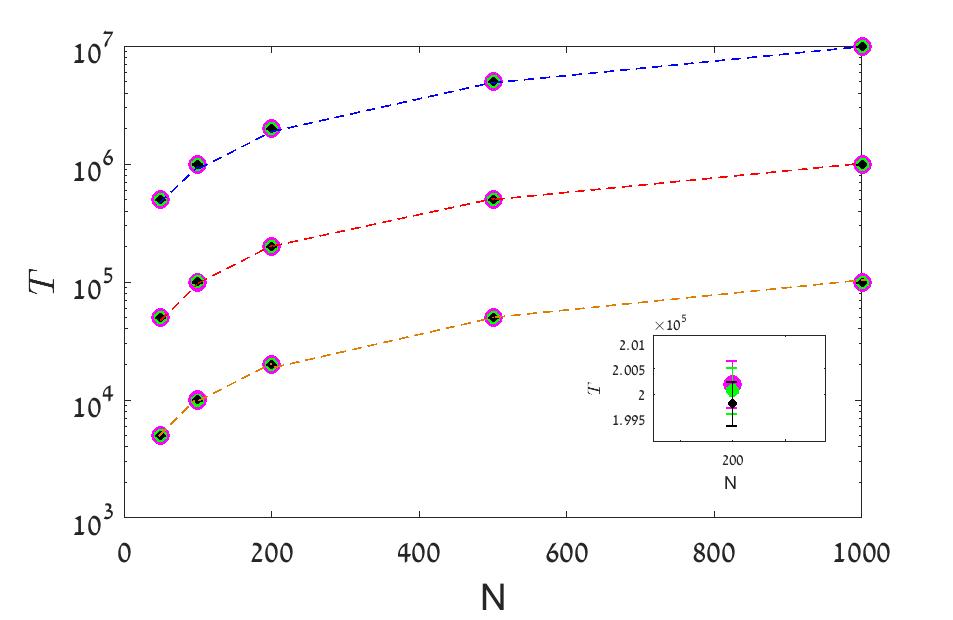}
\includegraphics[width=8cm]{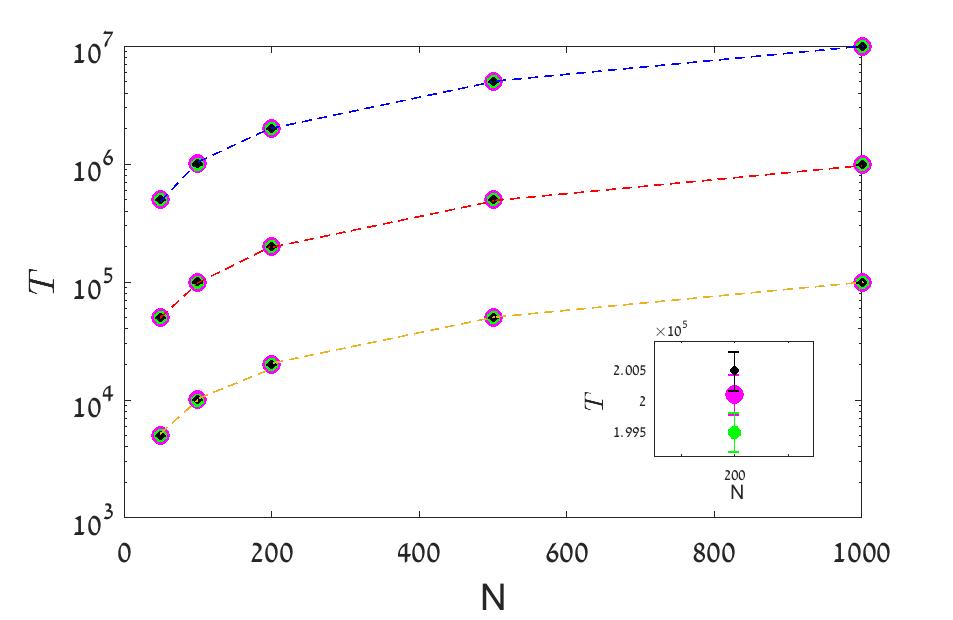}
\caption{Time to extinction $T$ (log scale) vs. $N$ for three different noise distributions. The mean (over 1000-2000 runs) time to extinction was measured as a function of $N=50,100,200,500,1000$, for $n_0=N$. The left panel present results for $\tau= \sigma = 0.1$ while in the right panel $\tau=\sigma=0.3$. For each $N$ and $\nu$ the value of $T$ is given for dichotomous noise (green circles), uniform distribution (magenta) and Beta distribution (black). Markers were chosen with different size to improve the visibility of the results. Dashed line were added manually to guide the eye and they connect results with $\nu = 0.01$ (yellow) $\nu=0.001$ (red) and $\nu = 0.0001$ (blue). In the insets the three points  at $N=200$, $\nu = 0.001$, with one standard deviation error bars, were magnified.  These error bars are too small and cannot be seen in the main panels.   }\label{simu}
\end{figure}

 To demonstrate the ability of a dichotomous noise to emulate the effect of other types of noise, we present in Figure \ref{simu} the outcomes of a few numerical experiments. The figures show the mean time to extinction vs. $N$ for our two-species competition model with one sided mutation, as described in the main text [Eq. (2)].  Three types of noise are compared.
 \begin{enumerate}
   \item $s(t)$ is either $\sigma$ or $-\sigma$ (dichotomous noise).
   \item $s(t)$ is picked from a uniform distribution between $(-\sigma \sqrt{12})$ and $(+\sigma \sqrt{12})$.
   \item $s(t)$  is picked from a beta distribution, $ \rm{Beta}(2,2)\sigma/\sqrt{0.05}$.
 \end{enumerate}
 All three distribution have a compact support, zero mean and variance $\sigma^2$.

 \clearpage

\section{Probability distribution function}

In the main text we have calculated the mean time to extinction, $T$, in the various phases of the logistic system. Here we would like present a few considerations regarding the full probability distribution function for extinction at $t$, $f(t) dt$ or the survival probability $Q(t) dt$. Of course $f(t) = -dQ(t)/dt$.

The state of our system is fully characterized by $P_{e,n}(t)$, the chance that the system admits $n$ A particles  at $t$, when the environmental state is $e$ (for dichotomous noise $e$ take two values that correspond to $\pm \sigma$). After a single birth-death event (time incremented from $t$ to $t+1/N$), the new state is given by
\begin{equation}
P_{e,n}^{t+1/N} = {\cal M } P_{e',m}^t,
\end{equation}
where ${\cal M }$ is the corresponding Markov matrix, ${\cal M }_{e,n;e',m}$ is the chance to jump from $m$ particles in environment $e'$ to $n$ particles in environment $e$.

The highest eigenvalue of the Markov matrix; $\Gamma_0=1$, corresponds to the extinction state, i.e, to the right eigenvector $P_{e,n} = \delta_{n,0}$ (at extinction  the state of the  environment insignificant) or the left eigenvector $(1,1,1,...)$. Using a complete set of left and right eigenvectors of this kind one may write $P_{e,n}(t)$ as,
\begin{equation} \label{sum1}
P_{e,n}(t) = \sum_k a_k v_k (\Gamma_k)^{Nt}.
\end{equation}
Here the index $k$ runs over all eigenstates of the Markov matrix, $v_k$ is the $k$-s right eigenvector, $a_k$ is the projection of $P_{e,n}(t=0)$ on the $k$-s left eigenvector and $Nt$ is the number of elementary birth-death events at time $t$ (for $t=1$, i.e., a generation, $Nt=N$). Writing $\Gamma_k = |\Gamma_k| \exp(\phi_k)$, one realizes that each $k$ mode decays like $\exp(-Nt \epsilon_k)$, when $\epsilon_k \equiv -\ln |\Gamma_k|$. Since the Markov matrix is real, eigenvalues are coming in complex conjugate pairs so $P_n$ is kept real and non negative at any time. For the extinction mode $\epsilon_0 =0$, all other modes have  $ \epsilon_k > 0$

Clearly, for any finite system the subdominant mode $\epsilon_1$ determines the maximal persistence time of the system, so at timescales above $t= 1/N \epsilon_1$ the chance of the system to survive, $Q(t)$, decays exponentially with $t$.

\begin{figure}
\includegraphics[width=5cm]{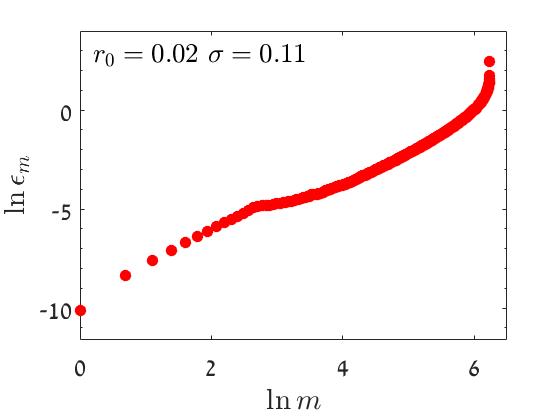}
\includegraphics[width=5cm]{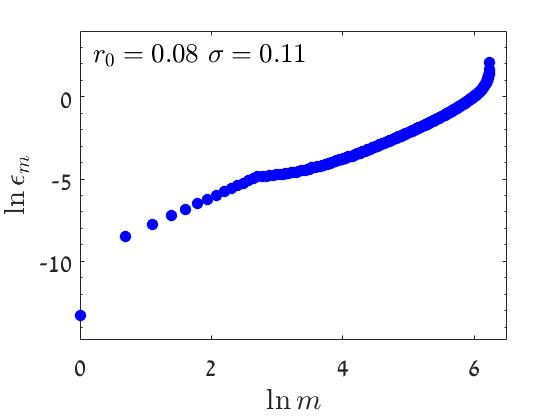}
\includegraphics[width=5cm]{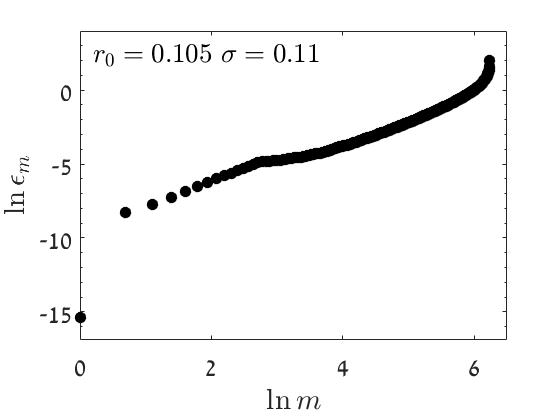}
\caption{The logarithm of the absolute value of the eigenvalues of the Markov matrix, $\epsilon_m$, is plotted against $\ln m$ for small $r_0$ (left panel), intermediate $r_0$ (middle panel) and large $r_0$ (right panel). The state with $m=1$ ($\ln m =0$) is the most persistent  non-extinction state. Clearly, as $r_0$ increases, a gap is opened between $\epsilon_1$ and $\epsilon_2$ (see figure \ref{fig3}). For $m>1$, the low-lying states satisfy $\epsilon_m \sim m^\rho$, where $\rho \approx 1.7$. Parameters are $\tau=1$, $\nu = 0.1$ and $N = 2^8$.  }\label{fig1s}
\end{figure}

\begin{figure}
\includegraphics[width=5cm]{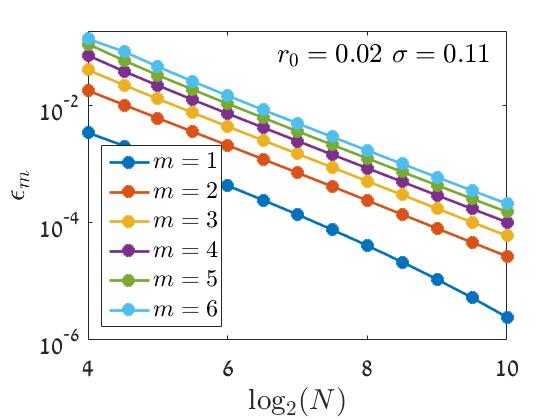}
\includegraphics[width=5cm]{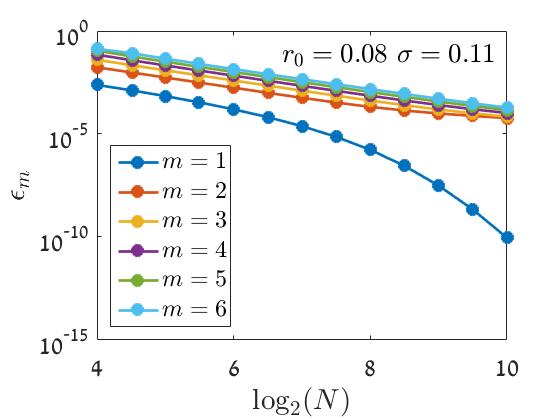}
\includegraphics[width=5cm]{beggriffithnew}
\caption{ $\epsilon_m$ (from $m=1$ to $m=6$, see legends) is plotted against $\log_2 N$ for different values of $r_0$.  Parameters are $\tau=1$ and $\nu = 0.1$.  }\label{fig2s}
\end{figure}

\begin{figure}[h]
\includegraphics[width=7cm]{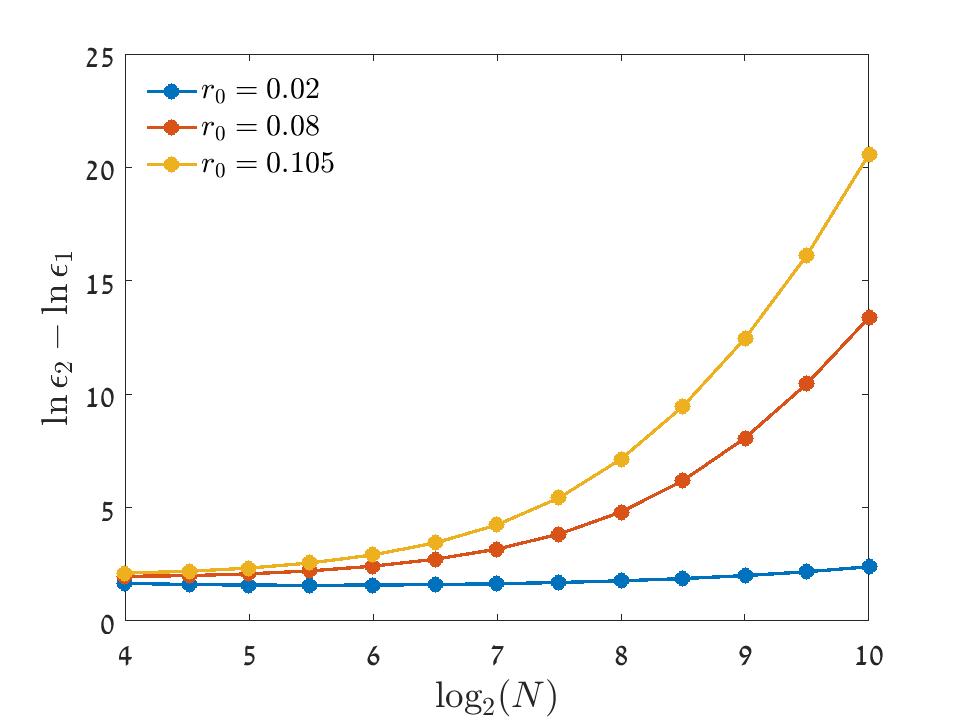}
\caption{ The gap, $\ln \epsilon_1- \ln \epsilon_2$, as a function of $ \log_2 N$. As $N$ increases the gap grows when $r_0$ is large  or intermediate but remains more or less fixed when $r_0$ is small. Parameters are $\tau=1$ and $\nu = 0.1$.   }\label{fig3s}
\end{figure}

Now one would like to make a distinction between two different situations. In the first, there is a \emph{gap} between $\epsilon_1$ and $\epsilon_2$, so when $N \to \infty$  $\epsilon_1  \ll \epsilon_2 $. This behavior is demonstrated in the right panels of Figures \ref{fig1s} and \ref{fig2s} and in Figure \ref{fig3s}.  In such a case the large $t$ behavior of the system is simply \begin{equation} \label{expo}
Q(t) dt = exp(-t/t_0),
 \end{equation}
where $t_0 \equiv 1/N \epsilon_1$. Accordingly, $f(t) = -\dot{Q} =  exp(-t/t_0)/t_0$ and the mean time to extinction, calculated in the main text, is $T = t_0 = 1/N \epsilon_1$. As showed in the main text, when $r_0>0$ $T$ grows with $N$, either exponentially or like a power law.

In the exponential phase the situation corresponds to this gap scenario, as discussed in~\cite{kessler2007extinction}. The purely exponential distribution (\ref{expo})  reflects an absence of memory: the system sticks for long times to its quasi stationary state $v_1$, and decay to zero on much shorter timescale due to rare events. This behavior is pronounced in Fig. 5 of \cite{ben2012coherence}. The decline to extinction may be a result of a  rare demographic event, like an improbable series of individual death, or  the result of an environmental rare event - an improbable series of bad years. In both cases, the \emph{decline time} (as defined in \cite{lande2003stochastic}) is short (logarithmic in $N$), so the exponential distribution reflects the accumulated chance of rare, short, and independent catastrophes.

The second scenario (demonstrated in the left panel of Figs \ref{fig1s} and \ref{fig2s} and in the blue line of Fig \ref{fig3s}) correspond to a \emph{gapless} system. Here the eigenvalues of ${\cal M}$ satisfy $\epsilon_m \sim  \epsilon_1 +c_1 (m-1)^\rho$, where $c_1$ is some tiny constant. In such a case the $\exp(- t N \epsilon_1 )$ factors out of the sum (\ref{sum1}), and the rest of the sum may be approximated by $\int exp(-c_1 t N m^{\rho}) dm$, yielding a power-law decay so,
\begin{equation} \label{lifetime}
Q(t) dt  \sim \frac{e^{-t/t_0}}{t^{1/\rho}} dt.
\end{equation}
In that case the mean time to extinction is not exactly  $t_0$ but the difference is only a numerical factor. If $\rho>1$ then,
\begin{equation}
T = t_0 (1-1/\rho),
\end{equation}
while if $\rho<1$ the ratio between $T$ and $t_0$ depends on the short time cutoff that must be imposed on the distribution (\ref{lifetime}) to avoid divergence at zero.

Now the decay is not purely exponential, since the system has long-term memory. Rare catastrophic events put an upper bound on the lifetime of the population, but extinction may occur, with relatively high probability, due to the random walk of the population size along the log-abundance axis.

\clearpage

\section{Population viability data}

\begin{figure}[h]
\includegraphics[width=8cm]{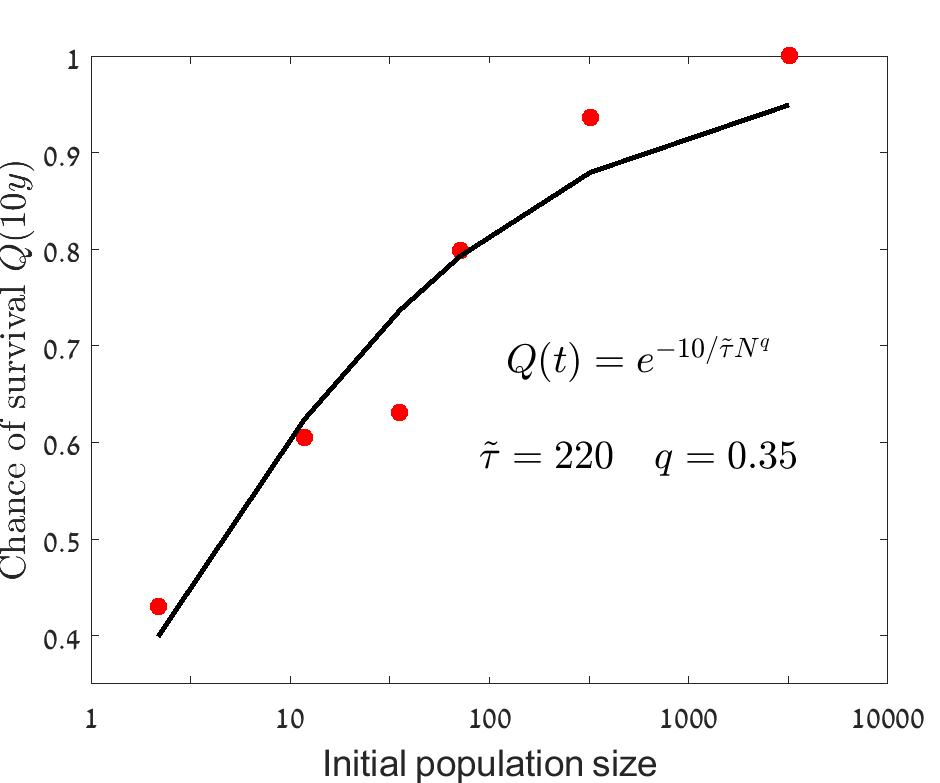}
\includegraphics[width=8.5cm]{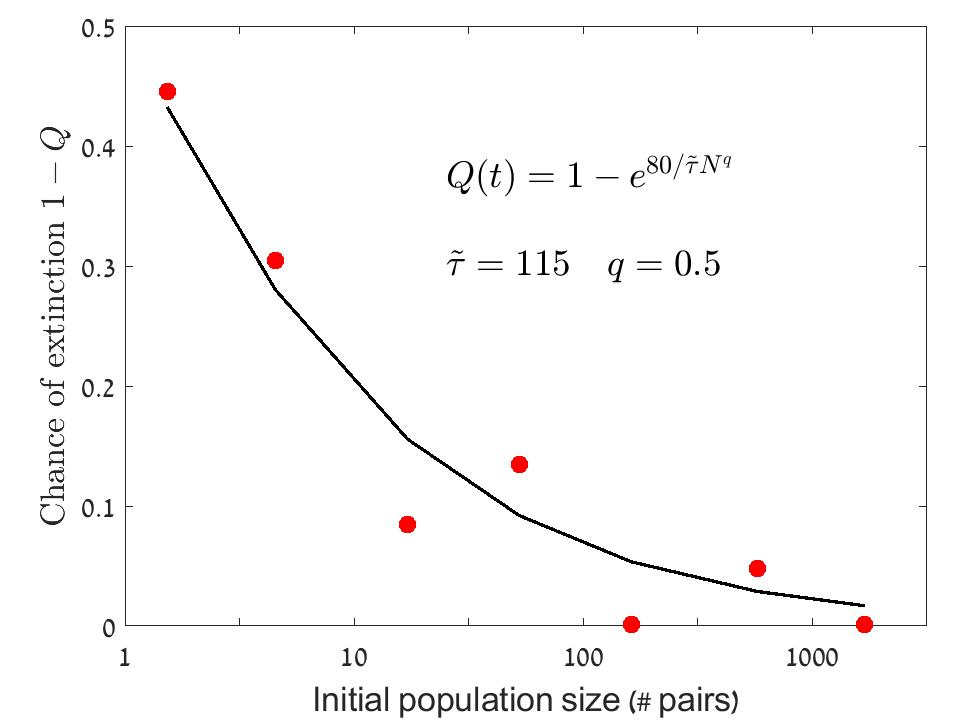}
\caption{The left panel (Figure 1 of \cite{matthies2004population}) shows the relationships between the size of plant populations in 1986 and their chance to survive 10 years later (red circles). The black line is the best fit to $Q(t)$, assuming that the mean time to extinction $T$ growth like  $N^q$. In the right panel we retrieved Figure 5 of \cite{jones1976short}, and the red circles correspond to  the chance of extinction of birds populations vs. the initial number of pairs (the last point in the original figure, that was too close to zero to be digitised, was omitted). The black line is the best fit to $1-Q(t)$, assuming that the mean time to extinction $T$ growth like  $N^q$.  }\label{birds}
\end{figure}

In the main text we have mentioned the population viability analysis of~~\cite{matthies2004population,jones1976short}. In Figure \ref{birds} we reproduce the relevant datasets from these two papers.

As one can see, both datasets (which are, of course, quite noisy because of the small number of samples in each bin, especially for the high abundance bins) allow for reasonable fits if the chance of survival, $Q(t)$, satisfies
\begin{equation}
Q(t) = \exp(-t/\tilde{\tau}N^q),
\end{equation}
 which is the expression one expects if the system is in the temporal Griffith phase. Note that the distinction between soft and sharp decline is irrelevant here, since the time window is fixed and we are interested only in the $N$ dependence.

When we tried to fit the data with  $Q = \exp(t/\tilde{\tau}\exp(\alpha N))$, as expected in the exponential phase, we ran into difficulties. In such a case one expects a much steeper dependence of $Q$ on $N$: if $T \sim \exp(\alpha N)$ than a chance in $N$ from $0.1/\alpha$ to $10/\alpha$, say,  takes $Q$ from vanishing values to one, so the survival probability  is a sharp sigmoid unless $\alpha$ takes very small values. As a result, for the plants 10y data our Matlab cftool  fit suggested an extremely tiny coefficient $\alpha = 0.01$, while for the birds 80y data it simply neglected the last four points and suggested $1-Q$ that drops to zero after the third point.

Moreover, both studies did not report a significant abundance decline in the surviving populations - in most of them abundance either grew up or kept fixed, see Figure 4 of~\cite{jones1976short} and Figure 4 of \cite{matthies2004population}. This implies that both systems are not in the logarithmic phase, where one should expect a general decrease in abundance for all populations.

We conclude that the most reasonable interpretation of the observed data is that the surveyed bird and plant populations are in the temporal Griffith phase, where the lifetime of a population scales $N^q$.

\end{document}